\documentclass[preprint2]{aastex}

\shorttitle{Autoclassification of the Variable 3XMM Sources}
\shortauthors{Farrell et al.}

\begin{document}


\title{Autoclassification of the Variable 3XMM Sources Using the Random Forest Machine Learning Algorithm}


\author{Sean A. Farrell\altaffilmark{1} and Tara Murphy\altaffilmark{1}}
\affil{Sydney Institute for Astronomy, School of Physics, The University of Sydney, Sydney, NSW, 2006, Australia}
\email{s.farrell@physics.usyd.edu.au}

\and

\author{Kitty K. Lo\altaffilmark{1}}
\affil{University College London, London, UK}


\altaffiltext{1}{ARC Centre of Excellence for All-Sky Astrophysics (CAASTRO)}


\begin{abstract}
In the current era of large surveys and massive data sets, autoclassification of astrophysical sources using intelligent algorithms is becoming increasingly important. In this paper we present the catalog of variable sources in the Third \emph{XMM-Newton} Serendipitous Source catalog (3XMM) autoclassified using the Random Forest machine learning algorithm. We used a sample of manually classified variable sources from the second data release of the \emph{XMM-Newton} catalogs (2XMMi-DR2) to train the classifier, obtaining an accuracy of $\sim$92$\%$. We also evaluated the effectiveness of identifying spurious detections using a sample of spurious sources, achieving an accuracy of $\sim$95$\%$. Manual investigation of a random sample of classified sources confirmed these accuracy levels and showed that the Random Forest machine learning algorithm is highly effective at automatically classifying 3XMM sources. Here we present the catalog of classified 3XMM variable sources. We also present three previously unidentified unusual sources that were flagged as outlier sources by the algorithm: a new candidate supergiant fast X-ray transient, a 400 s X-ray pulsar, and an eclipsing 5 hr binary system coincident with a known Cepheid. 
\end{abstract}





\section{Introduction}

Observational astronomy has entered a new era of large surveys that will produce incredible amounts of data at rates that are pushing beyond the limits of our ability to process in real time. Coinciding with this flood is an ever growing mountain of archival data that is increasingly under-utilised. Intelligent methods to quickly and accurately identify astrophysical sources are needed, with machine learning algorithms proving to be very effective in this respect. 

The Random Forest machine learning algorithm (hereafter referred to as RF) has shown great promise in the automatic classification of variable stars \citep{ric11,dub11}, the photometric classification of supernovae \citep{car10}, and most recently the automatic classification of variable X-ray sources \citep{lo14}. RF is an ensemble supervised classification algorithm that builds a forest of decision trees using a bootstrap sample from a training set of sources with known classification \citep{bre01}. It is one of the most accurate classification algorithms available \citep{car06}, is extremely fast, and can handle large data sets with a large number of features \footnote{In machine learning a feature is a measurable property of the object being classified, either a real number or a categorical label.}. In addition, there are only two parameters that the user needs to specify\footnote{There are more parameters that can be specified (e.g. it is possible to prune the trees, stop splitting once a particular node size is reached, or require that a minimum number of sources must be present in any given leaf), however only the two parameters described here are required. In this work we grew the trees fully without pruning.} -- the number of randomly selected features used at each node within the decision tree, and the number of trees in the forest -- making it extremely easy to use.

In previous work \citep{lo14} we investigated the feasibility of using RF to automatically classify the variable X-ray sources in the second data release of the Second \emph{XMM-Newton} Serendipitous Source Catalog \citep[2XMMi-DR2;][]{wat09}. At the time of its release in August 2008, 2XMMi-DR2 was the largest X-ray source catalog ever produced. 
%It contains 289,083 detections of 221,012 unique X-ray sources, of which 2,267 were flagged as variable. 
%Variable X-ray emission is produced by some of the most extreme physical phenomena in the Universe including the strongest magnetic fields, the highest temperatures, and extreme gravity. As such, variable X-ray sources are excellent probes of physical conditions that cannot be replicated in Earth-based laboratories.
We used the sample of 2XMMi-DR2 variable sources that had been manually classified (Farrell et al. in prep) as a training set for the RF classifier, obtaining an accuracy (evaluated through 10-fold cross-validation\footnote{The training set is divided into 10 sets. The model is then trained with nine sets and used to classify the remaining sample set, and then repeated for 10 different combinations. The accuracy is the total number of correctly classified samples divided by the total number of samples in the training set.}) of $\sim$97\%. \citet{lo14} also demonstrated the capability of RF to identify outlier sources that may represent rare new source populations, a stated scientific goal of most astronomical surveys. The training set was comprised of sources belonging to 7 categories: active galactic nuclei (AGN), cataclysmic variables (CVs), gamma ray bursts (GRBs), super soft sources (SSSs), stars, ultra luminous X-ray sources (ULXs), and X-ray binaries (XRBs).

The Third \emph{XMM-Newton} Serendipitous Source Catalog \citep[3XMM-DR4, hereafter referred to simply as 3XMM;][]{ros15} was released in July 2013 and contains 531,261 detections of 372,728 unique sources of which 3,696 are flagged as variable. 3XMM represents a $\sim$40\% increase in unique sources over 2XMMi-DR3, and a $\sim$63\% increase in variable sources over 2XMMi-DR2. 3XMM was constructed from 7,427 \emph{XMM-Newton} observations with the European Photon Imaging Cameras (EPIC) performed between 3 February 2000 and 8 December 2012, and is the largest X-ray source catalog so far released. 

In this paper we present a catalog of 3XMM variable sources that have been classified into six source categories using the RF classifier (hereafter referred to as the source class classification). As this is a serendipitous catalog, we would not expect a difference in the composition of sources in 2XMMi-DR2 versus 3XMM. We thus employed the same sample of 2XMMi-DR2 variable sources as used in \citet{lo14} for a training set. We also present the results of a study into the effectiveness of using the RF classifier to discriminate between spurious and real sources in the 3XMM catalog (hereafter referred to as the quality control classification). 
%In \S2 we describe the data preparation steps for our classification including the selection of features and the construction of training sets. In \S3 we present the classification methodology, including the construction of our training sets. The results of our classification are presented in \S4, along with our manual verification of the effectiveness of the technique. In \S5 we discuss a sample of previously unknown outlier sources that were identified by the classifier, and in \S6 we summarise the key points of the paper.

\section{Data Preparation \& Feature Selection}

Each unique source in both our training sets and our sample of unknown sources has multiple detections and thus a number of sets of X-ray features. In each observation a source may be detected by one or all of the three EPIC cameras, which in turn may have multiple exposures within a given observation (each with a unique light curve, although the other 3XMM features are the same for all exposures for a given camera within an observation). In addition, a number of fields were observed more than once providing additional detections taken at different epochs. We treated each detection independently in both the training and test sets and thus classified each detection separately. However, we combined the separate classifications from each detection to provide an overall classification (see \S3 for details of how this was performed).  

%For our classification we used X-ray properties taken directly from 3XMM, as well as features that were generated through a series of timing analyses applied to the light curves generated by the 3XMM pipeline. We also cross-matched our sample against radio, optical/near-infrared, and galaxy catalogs. 

With a few exceptions we used the same features for our classification as we used in \citet{lo14}. From 3XMM we took the four hardness ratios and errors, the Galactic latitude and longitude, the 0.2-12 keV (i.e. band 8) flux, the source extent (in arcseconds) and the (maximum) likelihood of the source being extended, the distance to the nearest neighbour in 3XMM, the source quality flag, and the confusion flag. The hardness ratios are defined as the ratio of count rates in two adjacent bands (normalised so as to always be between $-$1 and +1). They provide information on the crude shape of the X-ray spectrum and can thus be a powerful discriminator between different X-ray emissions mechanisms and thus different source types. However, when a source is not detected in either band used to calculate a hardness ratio, the resulting value is essentially a random number between $-$1 and +1. We therefore set the hardness ratio to a flag value of $-10$ and the error to 0 when both count rates used to calculate the hardness ratio were within 3$\sigma$ of zero. To generate the timing features, we analysed the 3XMM light curves using the same methods in \citet{lo14} after filtering out all points that lie outside the light curve good time intervals (GTIs). We searched for periodic variability \citep[using the generalised Lomb-Scargle periodogram from][]{zec09}, power law decays, flares \citep[using the Bayesian blocks technique;][]{sca98}, and also extracted a range of statistical features. A detailed description of how these features were extracted is given by \citet{lo14}. Table \ref{xray_features} provides a complete list of the X-ray and timing features included in our classification.

In addition to the X-ray features, we cross-matched our variable source sample against multi-wavelength catalogs. We used the Naval Observatory Merged Astrometric Dataset \citep[NOMAD;][]{zac04} for optical and near-infrared matches and the NRAO VLA Sky Survey \citep[NVSS;][]{con98}, the Sydney University Molonglo Sky Survey \citep[SUMSS;][]{mau03}, and the Second Epoch Molonglo Galactic Plane Survey \citep[MGPS-2;][]{mur07} to search for radio counterparts, using the 3$\sigma$ errors as a match criteria. When multiple counterparts were found we took the closest match as the correct one. Magnitudes in the BVR optical and JHK near-infrared bands, radio flux densities, as well B$-$V, V$-$R, J$-$H, and H$-$K colors were provided for those sources for which a counterpart was found. We also calculated X-ray to optical, X-ray to near-infrared, and X-ray to radio flux ratios for each band as well as the probability of a chance cross-match using the Bayesian method from \citet{bud08}. We note that due to the way that our training set sample was constructed (i.e. identifying 2XMM sources by matching against the SIMBAD and NED data bases) creates a bias towards brighter well known sources. As such a higher proportion of our training set sources have multi-wavelength matches than the overall 3XMM variable source sample, potentially leading the model to confuse fainter sources that do not have a match due to the limited sensitivity of the catalogs with brighter sources that by their nature do not have a multi-wavelength match. In an attempt to counteract this bias we set the multi wavelength features to flag values of $-1~\times~10^5$ for those sources where no counterpart was found, so that the model will down-weight the importance of the multi-wavelength properties for fainter sources\footnote{We note that there are numerous techniques for imputing values when data is missing from your sample. However, the lack of a multi-wavelength counterpart could either indicate that no counterpart is present (thus providing useful information about the nature of the source) or simply be due to the limited sensitivity of the multi-wavelength catalogs (which provides no information gain). We experimented with the imputation function provided in the \texttt{missForest} \citep{ste12} \texttt{R} library and found that while the overall classification accuracy did not vary significantly when missing data values were imputed, the accuracy for the minority GRB class dropped from 46$\%$ to 13$\%$ when imputation was employed.}. Similarly, for those sources where a counterpart was identified but magnitudes, fluxes, and/or colors were missing we set the relevant missing values to $-1~\times~10^5$. We chose flag values way outside the parameter space so that when SMOTE oversampling (see below) is employed the flag values will still remain significantly removed from the true parameter space.

In \citet{lo14} we cross-matched our sample against the Third Reference Catalog of galaxies \citep[RC3;][]{dev91} in order to identify which sources were potentially extragalactic. RC3 contains $\sim$23k galaxies within a distance $\lesssim$ 600 Mpc, with a mean distance of $\sim$40 Mpc and a standard deviation of $\sim$50 Mpc. For this work we instead used a sample of $\sim$1.4M galaxies $\lesssim$ 65 Gpc extracted from the NASA Extragalactic Database (NED) that had angular sizes and distances, which has a mean distance of $\sim$2 Gpc and a standard deviation of $\sim$4 Gpc. We cross-matched our sample against this NED galaxy catalog using the 3$\sigma$ X-ray source position and the galaxy D25 ellipse as a match criteria. When multiple matches were found we took the galaxy closest to the 3XMM position as the correct match. The features that were included for the galaxy cross-match include: a boolean flag indicating whether or not a match was found, the angular separation between the 3XMM source position and the galaxy centre, the ratio of the source/galaxy angular separation and the elliptical radius of the galaxy in the direction of the source ($\alpha$), and the log of the luminosity (calculated from the 0.2 -- 12 keV 3XMM flux and the galaxy distance). As for the multi-wavelength matches described above we set flag values for sources where no match was identified in order to counteract the bias towards brighter sources in our training sample. For sources where no match was found, we set the angular separation and $\alpha$ to 1 $\times$ 10$^5$ and log(L$_X$) to $-1$ as flags. Table \ref{mw_features} provides a complete list of the multi-wavelength and galaxy match features included in our classification.

\begin{table*}
\begin{center}
\caption{List of X-ray Features Used for Classification.\label{xray_features}}
\begin{tabular}{ll}
\tableline\tableline 
Feature & Description\\
\tableline
\multicolumn{2}{c}{X-ray Features}\\
\tableline
Inst & EPIC instrument (pn, MOS1, or MOS2) for each detection \\
HR1 &  Hardness ratio 1 (calculated from 0.2$-$0.5 keV and 0.5$-$1 keV count rates) \\
HR1$\_$err  & Error on hardness ratio 1\\
HR2 & Hardness ratio 2 (calculated from 0.5$-$1 keV and 1$-$2 keV count rates) \\
HR2$\_$err  & Error on hardness ratio 2\\
HR3 &  Hardness ratio 3 (calculated from 1$-$2 keV and 2$-$4.5 keV count rates)\\
HR3$\_$err  & Error on hardness ratio 3 \\
HR4 & Hardness ratio 4 (calculated from 2$-$4.5 keV and 4.5$-$12 keV count rates) \\
HR4$\_$err  & Error on hardness ratio 4\\
LII &  Galactic latitude (deg)\\
BII &  Galactic longitude (deg)\\
EP\_8\_FLUX  & Band 8 0.2$-$12 keV flux (erg cm$^{-2}$ s$^{-1}$)\\
EP\_EXTENT  & Source extent (arcsec)\\
EP\_EXTENT\_ML  & Maximum likelihood that the source is extended\\
DIST\_NN  & Distance to the nearest 3XMM source (arc sec)\\
SUM\_FLAG  & Source quality flag\\
CONFUSED  & Source confusion flag\\
\tableline
\multicolumn{2}{c}{X-ray Timing Features}\\
\tableline
%Flare finding
num\_flares   & Number of flares in the X-ray light curve \\
flare\_size1  &  Amplitude of the strongest flare (count s$^{-1}$)\\
flare\_time1   & Duration of the strongest flare (s) \\
%Lomb-Scargle periodogram
ls\_p1   & Period corresponding to the highest Lomb-Scargle periodogram peak (s)\\
ls\_p2   & Period corresponding to the 2nd highest Lomb-Scargle periodogram peak (s)\\
ls\_prob1   & False alarm probability of the highest Lomb-Scargle periodogram peak\\
ls\_prob2   & False alarm probability of the 2nd highest Lomb-Scargle periodogram peak\\
ls\_a1   & Amplitude of the most significant period in the Lomb-Scargle periodogram \\
ls\_a2   & Amplitude of the 2nd most significant period in the Lomb-Scargle periodogram\\
%Power law decay fit
A   & Inverse of the power law index for the power law model fit\\
F0   & Normalization of the best fit power law model\\
t0   & Time zero for power law decay fit \\
r\_chisq   & Reduced $\chi^2$ for the fit to the power law decay model\\
%Statistical tests
Amplitude   & 0.5 $\times$ [Max(rate) $-$ Min(rate)] (count s$^{-1}$)\\
Std   & Standard deviation of the X-ray light curve\\
Beyond1Std   & Percentage of data points in the light curve $>$ 1$\sigma$ from the weighted mean\\
Flux\_ratio\_mid20 & Ratio of the flux in the 60th to 40th percentiles over the 95th to 5th percentiles\\
Flux\_ratio\_mid35  & Ratio of the flux in the 67.5th to 32.5th percentiles over the 95th to 5th percentiles\\
Flux\_ratio\_mid50   & Ratio of the flux in the 75th to 25th percentiles over the 95th to 5th percentiles\\
Flux\_ratio\_mid65   & Ratio of the flux in the 82.5th to 17.5th percentiles over the 95th to 5th percentiles\\
Flux\_ratio\_mid80 &  Ratio of the flux in the 90th to 10th percentiles over the 95th to 5th percentiles\\
skew  & Skew of the distribution of count rates \\
Max\_slope   & Maximum slope of adjacent data points in the light curve (count s$^{-1}$) \\
Median\_abs\_dev  & Median of the absolute deviation from the mean count rate in the light curve\\
Med\_buffer\_range\_per   & Percentage of measurements within 20\% of the median\\
Percent\_amp   & Fractional difference between the highest count rate data point from the median\\
Per\_diff\_flux   & Difference between the 98th percentile and 2nd percentile count rates (count s$^{-1}$)\\
Mod\_index   & Variance/weighted mean\\
Fvar  &  Fractional rms variability of the X-ray light curve\\
\tableline
\end{tabular}
\end{center}
\end{table*}

\begin{table*}
\begin{center}
\caption{List of Multi-wavelength and Galaxy Match Features Used for Classification.\label{mw_features}}
\begin{tabular}{ll}
\tableline\tableline 
Feature & Description\\
\tableline
\multicolumn{2}{c}{Optical and Near-Infrared Features}\\
\tableline
B & NOMAD B-band magnitude\\
V & NOMAD V-band magnitude\\
R & NOMAD R-band magnitude\\
J & NOMAD J-band magnitude\\
H & NOMAD H-band magnitude\\
K & NOMAD K-band magnitude\\
B$-$V & Color calculated from NOMAD B and V-band magnitudes\\
V$-$R & Color calculated from NOMAD V and R-band magnitudes\\
J$-$H & Color calculated from NOMAD J and H-band magnitudes\\
H$-$K & Color calculated from NOMAD H and K-band magnitudes\\
Fx/Fb & Ratio of 0.2$-$12 keV X-ray to B-band flux\\
Fx/Fv & Ratio of 0.2$-$12 keV X-ray to V-band flux\\
Fx/Fr & Ratio of 0.2$-$12 keV X-ray to R-band flux\\
Fx/Fj & Ratio of 0.2$-$12 keV X-ray to J-band flux\\
Fx/Fh & Ratio of 0.2$-$12 keV X-ray to H-band flux\\
Fx/Fk & Ratio of 0.2$-$12 keV X-ray to K-band flux\\
nomad\_Bayes & Bayes factor for cross-match against NOMAD catalog\\
\tableline
\multicolumn{2}{c}{Radio Features}\\
\tableline
radio\_flux & Radio flux density from NVSS (1.4 GHz) or SUMSS/MGPS-2 (843 MHz) catalogs (mJy) \\
Fx/Frad & Ratio of 0.2$-$12 keV X-ray to radio flux\\
radio\_Bayes & Bayes factor for cross-match against radio catalogs\\
\tableline
\multicolumn{2}{c}{Galaxy Features}\\
\tableline
isGalMatch & Boolean flag indicating whether or not a match against a galaxy was found\\
galAngSep & Angular distance of 3XMM source from the galaxy centre (arcsec)\\
r\_ratio & Ratio of the angular distance to the galaxy centre over the radius of the galaxy ($\alpha$) \\
Luminosity & Log$_{10}$ of the 0.2$-$12 keV X-ray luminosity at the galaxy distance (log$_{10}$(erg s$^{-1}$))\\
\tableline
\end{tabular}
\end{center}
\end{table*}

\section{Methodology}

%In \citet{lo14} we demonstrated that RF is highly effective at classifying the unknown variable X-ray sources in 2XMMi-DR2. Here we apply the same technique to classifying the variable sources in the 3XMM catalog. In addition to our main goal of classifying sources into the various classes (e.g. AGN, CV, GRB etc.), we also investigated the application of RF to data quality control, i.e. identifying spurious source detections. 
%In the following section we describe the construction of the training sets for the classification of sources by sub-type (i.e. the main source classification) and for the data quality control classification. 

As in \citet{lo14} we used the \texttt{R} package \citep{R13} \texttt{randomForest} \citep{lia02} for our classification. This is an open source package that is freely available to the community, not proprietary code that we have written ourselves. To determine the optimal number of features we used the function \texttt{tuneRF} in \texttt{randomForest}, iterating through a range of values between 2 -- 20 for the number of features (with the number of trees set to 100) and comparing the out of bag (OOB) error for each run until a plateau was reached. We found the optimal number of features to be 15 for the source class classification and 16 for the quality control classification. To determine the optimal number of trees we again used \texttt{tuneRF} with the number of features set at the optimal value but varying the number of trees in each run again until a plateau in the OOB error was reached. For both classification runs the optimal number of trees was found to be 500.

To evaluate the accuracy of our classifiers we used the same 10-fold cross-validation method as outlined in \citet{lo14}. In our previous work we estimated an accuracy of $\sim$97\% for classifying the 2XMMi-DR2 variable sources \citep{lo14}. However, we used the entire training set for the 10-fold cross-validation, which unintentionally introduced a bias into the accuracy estimation\footnote{We note that this does not effect the classification of the unknown variable sources in 2XMMi-DR2 presented in \citet{lo14}, just the accuracy value reported.}. Our training and test sets were comprised of randomly selected detections for the cross-validation without cross-registration between detections by unique source number. As such, it is possible to have rows in both the training set and the sample to be classified that correspond to the same source and have almost identical features\footnote{A source detected in all three EPIC cameras in an observation will have at least three rows in the training set, all with identical multi-wavelength and galaxy match parameters and very similar (though not identical) X-ray features.}. If this happens the classifier is essentially classifying data in the test sample using a model built from (almost) the same data, thus producing an unrealistically high classification accuracy. To avoid this bias we randomly selected sources for the training and test sets for the cross-validation by unique source number, thus ensuring that all the detections for each unique source will either be in the training set or the test set, but never in both. However, we used the entire training set (i.e. with multiple rows per source per observation) to build the model for the classification of the unknown source sample. 

We expect that the higher the signal-to-noise (S/N) ratio of the X-ray detection the better RF should perform with respect to the classification of real sources, as the fractional error of the X-ray parameters will reduce with increased photon counts. We tested this assertion empirically, finding that the classification accuracy did indeed increase with the S/N of the X-ray detection (from 73$\%$ accuracy for detections with S/N $<$ 1 to 96$\%$ accuracy for detections with S/N $>$ 1000, evaluated through 10-fold cross-validation). To obtain the overall classification for each unique source in our main classification, we thus took the mean of the individual detection classifications for each source class weighted by the number of photon counts in that detection. The overall classification of a unique source is taken as the source class with the highest probability. For our quality control classification we simply took the mean of the individual detection classifications, as higher photon counts do not correspond with a more precise classification as the majority of spurious detections are due to the presence of very bright nearby sources and therefore have high S/N ratios.

\subsection{Training Set Construction}

\subsubsection{Source Class Classification}

For the source class classification, we used the same sample of manually classified variable 2XMMi-DR2 sources as in \citet{lo14}. The release of the 3XMM catalog involved a bulk reprocessing of all the \emph{XMM-Newton} data and thus includes a number of improvements that have been incorporated into the pipeline (such as improved source characterisation, astrometry, and greater sensitivity). We therefore cross-matched the \citet{lo14} training set of 873 2XMMi-DR2 sources against 3XMM in order to take advantage of these improvements, finding a match for 869 of them\footnote{Due to the improvements to the pipeline some sources present in previous versions of the catalog are not present in 3XMM or have shifted astrometry such that they do not match with 3XMM sources.}. 

In \citet{lo14} we separated the training set into 7 classes: AGN, CVs, GRBs, SSSs, STARs, ULXs, and XRBs. These classes made up the main types of sources identified through manual classification, and generally have very different physical properties. Although SSSs are a sub-class of CVs (with a white dwarf accreting from a binary companion undergoing steady thermonuclear burning), they have extremely soft X-ray spectra with very little emission above 1 keV and thus appear quite different to the bulk of other CVs, leading us to consider them as a different class of object in  \citet{lo14}. However, some novae (another sub-class of CVs present in our sample) have been observed to transition into a super-soft phase where they show similar X-ray properties to SSSs, but at other times look very different. We therefore combined our SSS and CV samples for this classification, as in principle there is no way that the classifier should be able to differentiate between a persistent SSS and a nova in a super-soft phase\footnote{In \citet{lo14} the classifier proved highly effective at discriminating between SSSs and CVs, despite there being a number of super-soft novae in the training sample. However, the majority of the SSSs in the sample were extragalactic while all but one of the CVs were in our own Galaxy. The classifier thus incorrectly placed significant weight on the galaxy match features.}. Indeed, without multiple observations of the same source over a long timescale (which would allow us to identify novae passing through a transient SSS phase and a persistent SSS) we are unable ourselves to discriminate between the two types of object. Table \ref{training_set} shows the breakdown of the training set into our 6 source classes.

As can be seen in Table \ref{training_set} our training set is heavily unbalanced, with the number of detections of the most abundant class (stars) outnumbering the rarest class (GRBs) by a factor of $\sim$240. This imbalance will significantly bias the model towards classifying an unknown object as the majority class (i.e. stars), leading to a higher accuracy for classifying stars but a lower accuracy for classifying the minority classes (in particular GRBs), despite rare objects being of particular interest to us. To compensate for this bias we oversampled all classes except for the stars using the SMOTE algorithm \citep{cha02}, which creates synthetic minority class samples with feature values selected using the k-nearest neighbours method from within the parameter space of the real sources belonging to a given class\footnote{We note that the use of SMOTE oversampling with data where flag values are employed to identify missing data (such as multi-wavelength and galaxy matches) may have unintended consequences as the flags will skew the parameter space to unphysical values. However, by selecting flags well outside the parameter space we attempted to force the model to produce values that are much closer to the flag values than real values. To test this we evaluated the classification accuracy both with and without SMOTE oversampling. We found that the RF model built with SMOTE and with flags representing missing data performed significantly better than a model without SMOTE and with imputed values for missing data when classifying the minority GRB class, and achieved similar accuracies when classifying the majority STAR class.}. We used the SMOTE implementation in the \texttt{DMwR} package \citep{tor10} in \texttt{R} to oversample the AGN by a factor of 3.5, the CVs by a factor of 7, the GRBs by a factor of 200, the ULXs by a factor of 15, and the XRBs by a factor of 2\footnote{We note that the total number of detections, not the number of unique sources, is oversampled.}. This oversampling thus provided approximately similar detection numbers for each class in the model as the dominant class of stars. To evaluate the impact of oversampling on the classification accuracies we built RF models both with and without SMOTE oversampling. We found that the overall 10-fold cross-validated classification accuracy and the accuracy for classifying the majority star class did not change significantly when oversampling was employed (they were consistent with the non-oversampled accuracies within 1$\%$). However, the classification accuracy for the minority GRB class improved significantly (from 13$\%$ to 46$\%$) with oversampling. As such, we chose to use oversampling for our subsequent analyses.

\begin{table}
\begin{center}
\caption{Number of Sources and Classification Accuracy for the Training Sets and Unknown Sample\label{training_set}}
\begin{tabular}{lccc}
\tableline\tableline 
Class & Sources & Detections & Accuracy$^a$\\
\tableline
AGN & 99 & 435 & 90\% \\
CV & 91 & 219 & 75\% \\
GRB & 8 & 8 & 63\% \\
STAR & 571 & 1,931 & 99\% \\
ULX & 17 & 110 & 59\% \\
XRB & 83 & 632 & 77\% \\
\tableline
%Total & 869 & 3,335\\
REAL & 867 & 8,841 & 98\% \\
SPURIOUS & 363 & 1,939 & 89\% \\
\tableline
Unknowns & 2,876 & 18,619 & \nodata \\
\tableline
\multicolumn{4}{l}{$^a$Overall classification accuracy for each class}\\
\multicolumn{4}{l}{from 10-fold cross-validation.}\\
%$^a$Overall classification accuracy for each class from 10-fold cross-validation.
\end{tabular} 
\end{center}
\end{table}

\subsubsection{Quality Control Classification}

Despite the incorporation of significant improvements in the PSF modelling \citep{rea11}, the 3XMM pipeline source detection algorithm still occasionally detects spurious point sources around bright sources, in crowded fields, and in diffuse emission. Spurious detections can also occur due to optical loading from bright stars, as the EPIC cameras (particularly the MOS detectors) are sensitive to bright optical emission that can produce features in the images that are incorrectly detected as X-ray sources. Each source in the catalog is automatically assigned a quality flag (SUM\_FLAG) by the pipeline based upon its proximity to regions that may cause issues with the reliability of source parameters/products or where spurious detections commonly occur (e.g. near a bright point source or diffuse emission). 

While filtering out sources with quality flags $\geqslant$ 2 will provide a reliable sample, the way in which the flags are assigned means that such a filtering criteria will also discard bright real sources (around which spurious detections are common) that are potentially of interest to the user. 

Spurious detections are easily identifiable through inspection of images and source products, so we hypothesised that spurious sources could be automatically identified via RF classification. To test this, we constructed a binary training set of non-spurious and spurious sources. For the sample of non-spurious sources we used the sample of spurious sources that were identified during our manual classification of the 2,267 2XMMi-DR2 variable sources (Farrell et al. in prep). For the sample of non-spurious sources we used the same sample described in \S3.1.1 that we used for our main classification (with the source class simply set to `REAL'). As with the main classifier training set, we first cross-matched our spurious and non-spurious sample of sources against 3XMM so as to obtain updated and improved source parameters. We used the same features for the classification as used for the source class classification. Table \ref{training_set} shows the breakdown between real and spurious sources in our training set.

Our quality control sample is less unbalanced than our main training set, yet the real detections sill outnumber the spurious detections significantly (Table \ref{training_set}). We thus oversampled the spurious source sample using the SMOTE algorithm \citep{cha02} by a factor of 4 to provide approximately the same number of detections as for the real class\footnote{Again, we note that we oversampled the total number of detections not the number of unique sources.}. 

\section{Classification Results \& Verification}

Table \ref{classified} shows the results of the classification of the 2,876 unknown 3XMM variable sources by source class and for quality control (full version available online).  As described in \S3, each detection of each unique source was classified separately and then the overall accuracy was calculated by combining the detection classifications. We obtained an overall accuracy of $\sim$92\% for our classification by source class, evaluated by 10-fold cross-validation. For the quality control classification, the overall accuracy was $\sim$95\%. When considering each detection separately the classification accuracies were lower, with $\sim$87\% for our classification by source class\footnote{For comparison, we re-ran the classification using the 2XMMi-DR2 training set in \citet{lo14} but with the training and test sets randomly selected based on the unique source ID (i.e. to ensure no cross-over between the test and training sets), and obtained an accuracy of $\sim$92\%.} and $\sim$94\% for our quality control classification.

\begin{figure}
\includegraphics[width=\columnwidth]{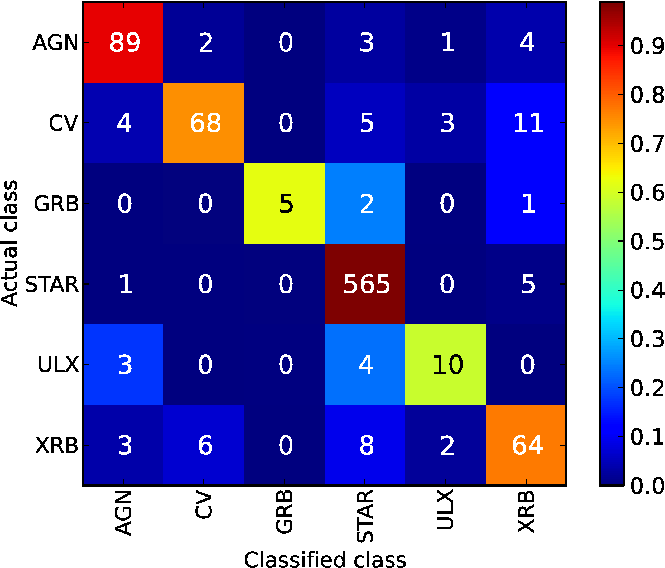}
\caption{Confusion matrix for the classification by source class.}\label{cm_real}
\end{figure}

\begin{figure}
\includegraphics[width=\columnwidth]{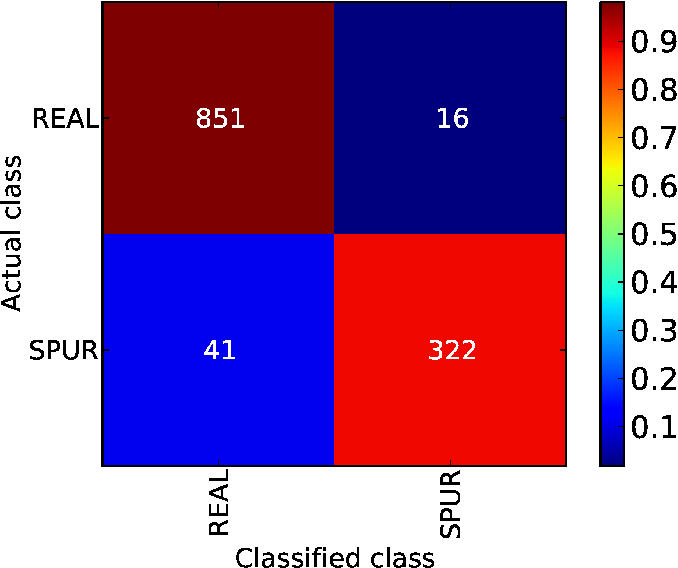}
\caption{Confusion matrix for the spurious vs non-spurious source classification.}\label{cm_spur}
\end{figure}

Figures \ref{cm_real} and \ref{cm_spur} show the confusion matrices for the classification by source class and the quality control classification, respectively. The number in each square in the confusion matrices represents the overall classification for each unique source compared to the actual classification obtained through manual inspection. The classification accuracy for each source class is given in Table \ref{training_set}, calculated as the number of correct source classifications over the total number of sources in each class of the training set.

\begin{table*}
\begin{center}
\caption{3XMM Variable Source Classifications.\label{classified}}
\begin{tabular}{lccccccccccc}
\tableline\tableline 
3XMM Name	&	P$_{AGN}$	&	P$_{CV}$	&	P$_{GRB}$	&	P$_{STAR}$	&	P$_{ULX}$	&	P$_{XRB}$	&	P$_{Max}$	&	P$_{Spur}$	&	Class	&	Outlier	&	Margin	\\
\tableline 
J000055.5+443710	&	0.000	&	0.002	&	0.000	&	0.994	&	0.004	&	0.000	&	0.994	&	0.002	&	STAR	&	0	&	0.987	\\
J000209.5-300035	&	0.122	&	0.143	&	0.000	&	0.555	&	0.113	&	0.067	&	0.555	&	0.010	&	STAR	&	364	&	0.109	\\
J000219.7-295607	&	0.001	&	0.003	&	0.000	&	0.996	&	0.000	&	0.000	&	0.996	&	0.000	&	STAR	&	2	&	0.991	\\
J000222.8-060559	&	0.014	&	0.081	&	0.018	&	0.369	&	0.043	&	0.474	&	0.474	&	0.005	&	XRB	&	28	&	-0.052	\\
J000300.6-294942	&	0.024	&	0.050	&	0.000	&	0.907	&	0.005	&	0.013	&	0.907	&	0.003	&	STAR	&	24	&	0.815	\\
J000334.5-295830	&	0.017	&	0.059	&	0.000	&	0.913	&	0.007	&	0.003	&	0.913	&	0.001	&	STAR	&	16	&	0.826	\\
J000354.2-255841	&	0.004	&	0.022	&	0.000	&	0.962	&	0.007	&	0.004	&	0.962	&	0.000	&	STAR	&	3	&	0.924	\\
J000511.8+634018	&	0.058	&	0.115	&	0.017	&	0.583	&	0.050	&	0.176	&	0.583	&	0.008	&	STAR	&	294	&	0.167	\\
J000532.8+200717	&	0.776	&	0.095	&	0.002	&	0.073	&	0.002	&	0.053	&	0.776	&	0.010	&	AGN	&	56	&	0.552	\\
J000612.2+201304	&	0.037	&	0.109	&	0.000	&	0.468	&	0.049	&	0.336	&	0.468	&	0.910	&	STAR	&	594	&	-0.064	\\
J000613.6+201118	&	0.017	&	0.073	&	0.011	&	0.243	&	0.084	&	0.572	&	0.572	&	0.963	&	XRB	&	50	&	0.143	\\
J000613.6+201253	&	0.044	&	0.131	&	0.002	&	0.343	&	0.012	&	0.468	&	0.468	&	0.918	&	XRB	&	32	&	-0.063	\\
J000618.2+201248	&	0.021	&	0.092	&	0.023	&	0.255	&	0.132	&	0.477	&	0.477	&	0.937	&	XRB	&	75	&	-0.045	\\
J000621.5+201149	&	0.025	&	0.136	&	0.026	&	0.259	&	0.118	&	0.435	&	0.435	&	0.922	&	XRB	&	63	&	-0.129	\\
J000627.0+200904	&	0.028	&	0.210	&	0.002	&	0.420	&	0.048	&	0.292	&	0.420	&	0.862	&	STAR	&	601	&	-0.160	\\
J000631.0+200720	&	0.031	&	0.204	&	0.000	&	0.412	&	0.087	&	0.265	&	0.412	&	0.985	&	STAR	&	428	&	-0.175	\\
J000634.7+200548	&	0.049	&	0.146	&	0.000	&	0.456	&	0.057	&	0.292	&	0.456	&	0.925	&	STAR	&	459	&	-0.087	\\
J000635.5+200527	&	0.047	&	0.201	&	0.000	&	0.421	&	0.071	&	0.260	&	0.421	&	0.961	&	STAR	&	786	&	-0.158	\\
J000638.9+200403	&	0.038	&	0.201	&	0.002	&	0.388	&	0.050	&	0.321	&	0.388	&	0.963	&	STAR	&	987	&	-0.225	\\
J000639.6+200343	&	0.046	&	0.187	&	0.002	&	0.455	&	0.059	&	0.251	&	0.455	&	0.973	&	STAR	&	715	&	-0.090	\\
\tableline 
\multicolumn{12}{l}{\textbf{Notes.} Column 1: 3XMM name. Columns 2-7: probability given by our RF classifier that the source belongs to}\\
\multicolumn{12}{l}{one of the training set source classes. Column 8: the maximum probability of the classification by source class.} \\
\multicolumn{12}{l}{Column 9: the probability given by our quality control RF classifier that the source is spurious (averaged over all} \\
\multicolumn{12}{l}{detections). Column 10: class given to the source by our classifier (calculated as the mean classifications over all} \\
\multicolumn{12}{l}{detections weighted by the number of photon counts in that detection). Column 11: the outlier measure of the} \\
\multicolumn{12}{l}{source. Equation 10 in \citep{lo14} provides a definition of this parameter. Larger values indicate a higher} \\
\multicolumn{12}{l}{likelihood of being an outlier. Column 12: the classification margin of the source (Margin = 2 $\times$ P$_{Max}$ $-$ 1).} \\
%\multicolumn{12}{l}{}\\
%\tableline
\end{tabular}
\end{center}
\end{table*}

% Confusion matrix plots

\begin{figure}
\includegraphics[width=\columnwidth]{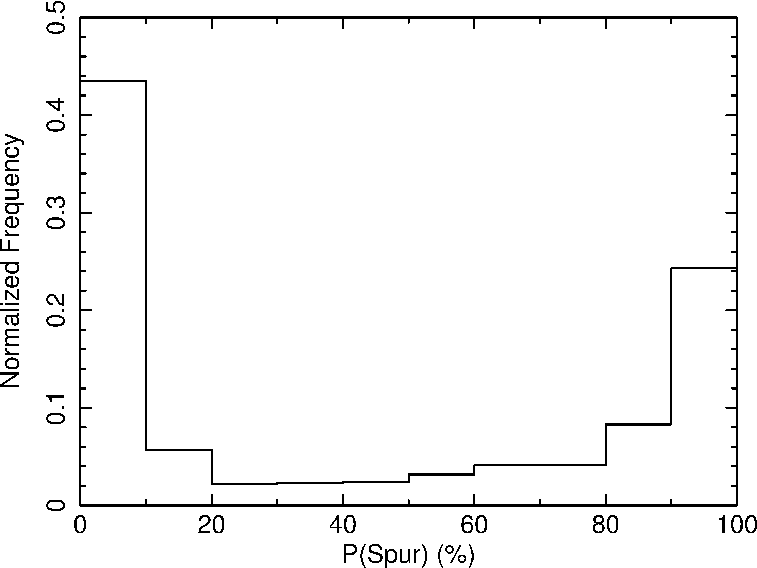}
\caption{Distribution of spurious probabilities for the classified unknown sample of 2,876 sources.}\label{spur}
\end{figure}

\begin{figure}
\includegraphics[width=\columnwidth]{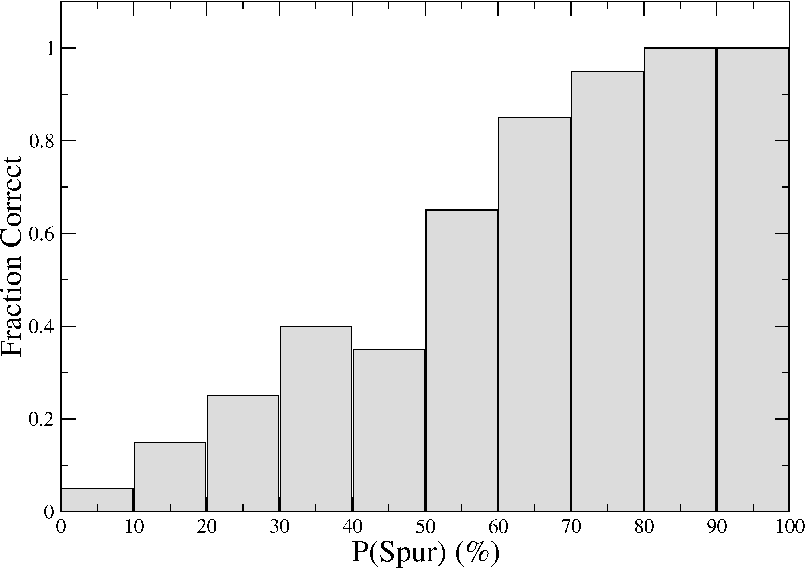}
\caption{Results of manual verification for quality control classification. The histograms represent the fraction of sources that were correctly classified by Random Forest as a function of the probability of being spurious. There were 20 sources evaluated per P(Spur) bin.}\label{mv_spur}
\end{figure}

The distribution of the probability of being spurious is bimodal, with $\sim$50\% of the unique sources having P(Spur) $\leqslant$ 20\% and $\sim$30\% having P(Spur) $\geqslant$ 80\% (see Figure \ref{spur}). In order to test the overall accuracy of the quality control classification, we randomly selected a sample of 200 sources and manually inspected their 3XMM products (i.e. images, light curves, spectra etc.) to determine whether or not they were spurious. Of these 200 sources, 96\% were correctly classified as either real (i.e. P(Spur) $<$ 50\%) or spurious (i.e. P(Spur) $\geqslant$ 50\%) by the algorithm, consistent with the accuracy of $\sim$95\%~obtained through 10-fold cross-validation. To investigate the accuracy as a function of P(Spur), we randomly selected 20 sources from each 10-percentile P(Spur) bin and performed the same manual verification (see Figure \ref{mv_spur}). As expected, a high P(Spur) corresponds to a high probability that a sources is spurious, while the majority of sources with low P(Spur) values are real. Taking P(Spur) $\leqslant$ 30\% should thus provide a reliable sample of sources. To test this, we randomly inspected 200 sources with P(Spur) $\leqslant$ 30\%, finding that 97\% were indeed real sources.

% Prob distribution plots

\begin{figure}
\includegraphics[width=\columnwidth]{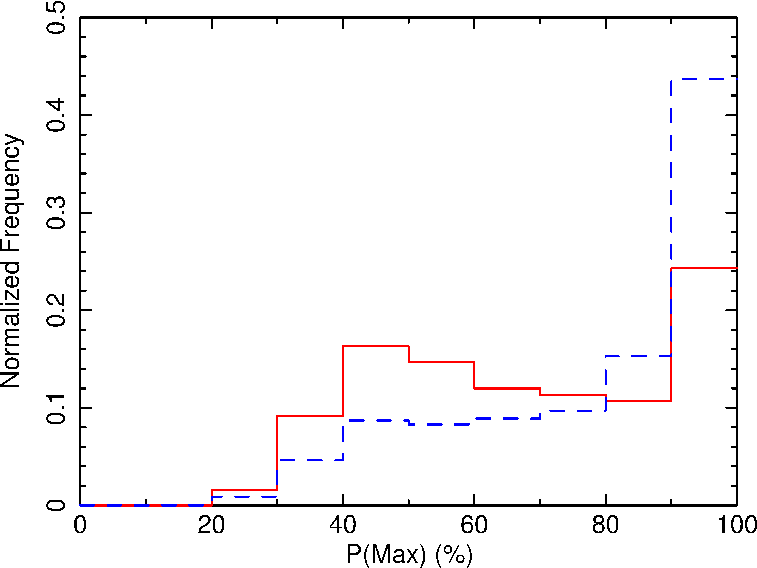}
\caption{Distribution of maximum classification probabilities for the classified unknown sample. The solid red histogram shows the distribution for the entire sample (2,876 sources), while the dashed blue histogram shows the distribution for the clean sub-sample (1,475 sources), i.e. sources with P(Spur) $\leqslant$ 0.3.}\label{pmax}
\end{figure}

The distribution of maximum probabilities for the classification by source class is also bimodal, with the maximum peak around P(Max) $\sim$ 95\% and a second broader peak around P(Max) $\sim$ 45\% (see Figure \ref{pmax}). Filtering out sources with P(Spur) $\leqslant$ 30\% primarily discards sources with low P(Max) values, indicating as expected that RF has trouble classifying spurious sources into real source classes. Table \ref{class_bdown} shows the breakdown of classified unknown sources by source class for the entire sample (`all') and for the sample with P(Spur) $\leqslant$ 30\% (`good'). To verify the accuracy of the classifier we manually inspected a random sample of sources with P(Spur) $\leqslant$ 30\%, checking for identifications within SIMBAD and NED. We identified 101 real sources that had an identification in the literature, of which 92\% were in agreement with the classification provided by RF, consistent with the accuracy of $\sim$92\%~obtained by 10-fold cross-validation.

\begin{table}
\begin{center}
\caption{Breakdown of Classified Sources\label{class_bdown}}
\begin{tabular}{lcccc}
\tableline\tableline 
Class & All & \% &Good &  \%\\
\tableline
AGN & 144 & 5.0\% & 107 & 7.3\% \\
CV & 152 & 5.3\% & 70 & 4.7\%\\
GRB & 5 & 0.2\% & 5 & 0.3\%\\
STAR & 1,942 & 67.5\% & 1,162 &  78.8\% \\
ULX & 54 & 1.9\% & 11 &  0.7\%\\
XRB & 579 & 20.1\% & 120 &  8.1\% \\
\tableline
Total & 2,876 & 100\% & 1,475 & 100\% \\
\tableline
\end{tabular}
\end{center}
\end{table}

\begin{figure}
\includegraphics[width=\columnwidth]{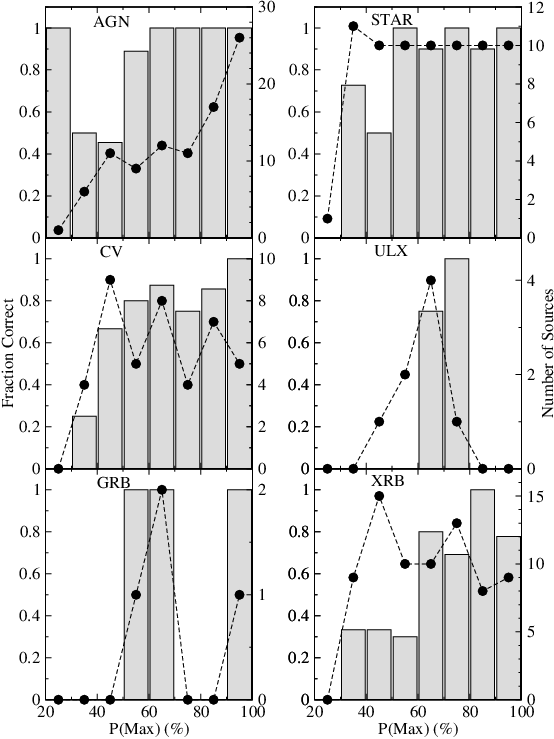}
\caption{Results of manual verification by source class. The histograms represent the fraction of sources that were correctly classified by Random Forest as a function of the maximum probability. The black circles indicate the number of sources in each P(Max) bin.}\label{mv_class}
\end{figure}

\begin{figure}
\includegraphics[width=\columnwidth]{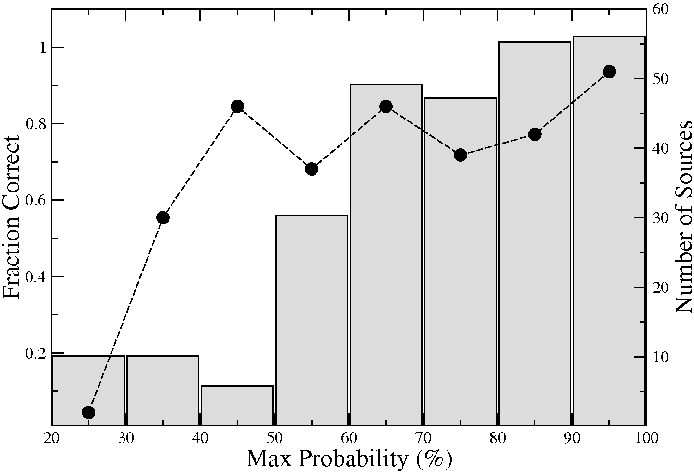}
\caption{Results of manual verification for overall classification. The histogram represents the fraction of sources that were correctly classified by Random Forest as a function of the maximum probability. The black circles indicate the number of sources in each P(Max) bin.}\label{mv_all}
\end{figure}

We next evaluated the classifier accuracy for each source class and as a function of P(Max), considering only the sample of `good' sources (i.e. P(Spur) $\leqslant$ 30\%). There were $\leqslant$ 120 good sources in each class except for the sources classified as stars (see Table \ref{class_bdown}). We thus manually checked all good sources that were classified as an AGN, CV, GRB, ULX, or XRB for identifications in SIMBAD and NED. For the stars, we checked a sample of sources randomly selected from each 10-percentile P(Max) probability bin between 40-100\% until we had found 10 sources in each bin with a literature identification. There were only 4 sources in the 20-30\% bin so we inspected all of them. In the 30-40\% bin there were 29 sources of which 11 had a match in SIMBAD or NED, all of which we included in our sample. The results of this manual verification are shown in Figure \ref{mv_class}. For the classes with decent sampling (i.e. the AGN, CV, STAR, and ULX classes) it is clear that (as expected) the classification accuracy is correlated with P(Max). This is demonstrated even more clearly in Figure \ref{mv_all}, which presents the accuracy per P(Max) bin across all source classes. Selecting classified sources with P(Spur) $\leqslant$ 30\% and P(Max) $\geqslant$ 60\% should thus provide a clean sample of real sources with correct classifications. 

The classifier performed particularly well on the minority GRB class. Five sources in the unknown sample were classified as GRBs, all of which were real sources with very low probabilities of being spurious as determined by the quality control classification. Four of these objects were known GRBs that were the targets of the observations. The fifth source (3XMM J054707.6+001742) has not previously been identified but demonstrates a power law decay in its light curve and has a spectrum that could be consistent with an absorbed moderately steep power law, similar to what is observed from known GRBs. However, the spectrum also shows very strong iron line emission which is not observed from other GRBs, and its location is consistent with a nebula rather than a galaxy thus indicating that it is most likely a star rather than a GRB. We also investigated those sources where the highest classification probability did not indicate a GRB, but where the GRB probability was the second highest assigned by the classifier. Twelve sources met this criterion, of which 5 were real sources. Two of these sources were known stars, while another was a known GRB (again, the target of the observation). The two remaining real sources have not previously been classified, but are likely to be stars due to their coincidence with bright point like optical sources.

\begin{figure*}
\includegraphics[width=\textwidth]{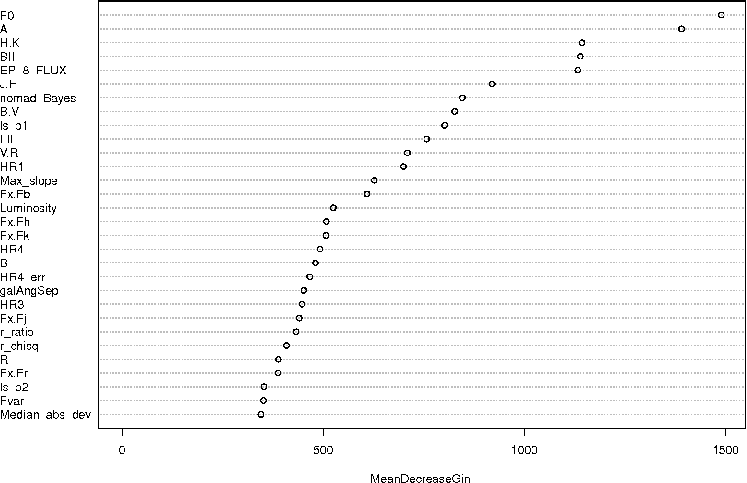}
\caption{Relative importance of the features for the classification by source class. The higher the value of the mean decrease Gini impurity, the higher the relative importance of a feature.}\label{gini_real}
\end{figure*}

In order to estimate the relative importance of each feature, we calculated the Gini index which measures the total decrease in node impurities from splitting on a given feature, averaged over all the trees in the forest \citep[see Equation 1 in][]{lo14}. Figure \ref{gini_real} shows the relative feature importance of the top 30 features for the classification by source type. Surprisingly, the normalization and slope of the power law model fitted to the light curves are the most important features, followed by the H$-$K near-infrared color, the Galactic latitude, and the X-ray flux. This differs from the \citet{lo14} results, which found the five most important features to be (in order of decreasing importance) the X-ray flux, X-ray luminosity, X-ray hardness ratio HR3, K-band magnitude, and the r\_ratio (i.e. $\alpha$). It is probable that this discrepancy is due to the significant increase in the number of detections in the training set used in this work, achieved as a result of including detections that did not have the light curve timing features. Overall, the optical and near-infrared features (specifically the colours and X-ray to optical/near-infrared flux ratios) appear to be highly informative, as do the X-ray flux and hardness ratios. 
Nonetheless, the inclusion of the timing features has a significant effect on the model accuracy as the 10-fold cross-validation accuracy drops to $\sim$85$\%$ when the timing features are removed.

\begin{figure*}
\includegraphics[width=\textwidth]{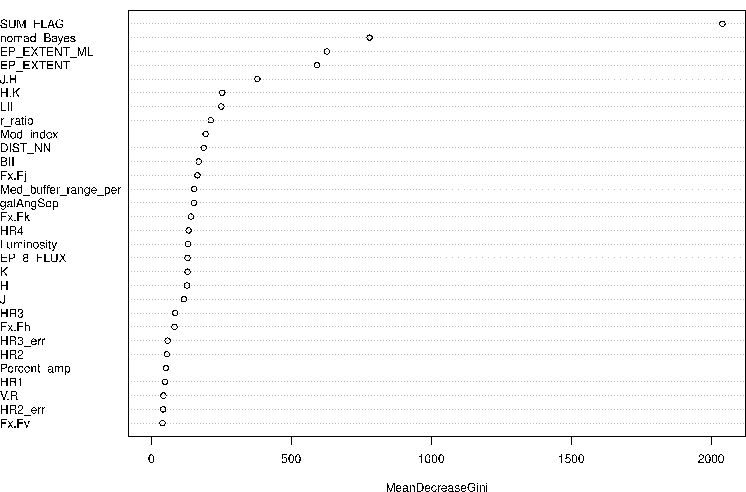}
\caption{Relative importance of the features for the quality control classification. The higher the value of the mean decrease Gini impurity, the higher the relative importance of a feature.}\label{gini_spur}
\end{figure*}

We also calculated the relative importance of the features for the quality control classification (see Figure \ref{gini_spur}). As expected, the 3XMM quality flag (SUM\_FLAG) is by far the most important feature, followed by the probability of a chance cross-match with a NOMAD source (i.e. the nomad\_Bayes feature), the EPIC extent maximum likelihood and extent, and the J$-$H color. Somewhat surprisingly, the distance to the nearest neighboring 3XMM source was not an important feature, and the 3XMM confusion flag did not rate in the top 30 feature list at all. This indicates that confusion with nearby sources is not a major issue with regards to spurious source detection.

\section{Outlier Sources}

In addition to classifying sources that belong to known classes, RF can also be used to identify sources that belong to novel classes that were not in the training set. While these outlier sources should have low P(Max) probabilities, this alone is insufficient to identify truly anomalous sources as missing information (e.g. a lack of mutli-wavelength or galaxy matches, poor S/N X-ray data etc.) will also produce low classification probabilities. A better method of identifying anomalous sources is to use the outlier measure, which represents the proximity of a given unknown source classified by RF to the training set source population for the same class. For each classified unknown source we calculated the proximity matrix and outlier measure using the \texttt{randomForest} package in \texttt{R} \citep[see Equation 10 in][]{lo14}. We also calculated the classification margin, which is the difference between the probability of the source belonging to the class with P(Max) and the probability that it does not belong to that class, i.e. Margin = 2 $\times$ P(Max) $-$ 1. Both the outlier measure and classification margin are provided for each of the classified unknown sources (see Table \ref{classified}). 

\begin{figure}
\includegraphics[width=\columnwidth]{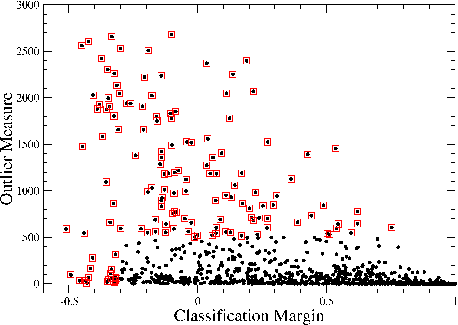}
\caption{Classification margin vs outlier measure for the good classified unknown sources. The red squares indicate the 144 sources that were investigated further.}\label{outliers}
\end{figure}

Figure \ref{outliers} shows the classification margin vs outlier measure for the good sample of classified unknown samples. We selected a sample of 144 good sources with a margin $\leqslant$ $-0.3$ or an outlier measure $\geqslant$ 400 for further investigation. Of these, 4 were found to be spurious detections, 30 sources were previously identified (20 of which were classified correctly by RF),  while the remaining 110 had no match in either SIMBAD or NED. The sample of previously identified sources contains a number of true rare sources that were not in the training set. These include: a soft X-ray transient, an isolated neutron star (one of the magnificent seven), a soft gamma repeater (magnetar), a semi-detached Beta Lyra eclipsing binary, two candidate intermediate mass black holes (identified previously through their unusual variability), and a highly unusual Seyfert 2 AGN with a $\sim$3.8 hr period and extremely soft X-ray spectrum \citep[possibly hosting an intermediate mass black hole;][]{ho12}. Also present in this sample is the source 3XMM J180658.7$-$500250, which was previously identified as an outlier in \citet{lo14} and is also thought to be an unusual type of AGN.

A number of sources were detected in mosaic mode observations (primarily of Jupiter or Mars, where the attitude was stepped during the observation) that are known to be problematic and have unreliable X-ray data and products. All sources in the affected observations appear to show the same highly unusual variability and are also likely to have unreliable astrometry, leading to issues with the cross-matching against multi-wavelength and galaxy catalogs. This combination makes them outliers compared to the training set, though in this case due to data processing issues rather than the nature of the sources. 

Of the remaining known sources, 16 were classified as stars and had low numbers of X-ray photon counts, leading to low S/N data and large scatter in their X-ray properties.  Five sources were previously identified AGN, one was a known nova (possibly in a super-soft phase), and another was a known ULX. Many of the stars appear to have high proper motions such that cross-matching against the NOMAD catalog either found no counterpart or an incorrect match. Three of the outlier sources (all of which were classified as stars) had good S/N and showed truly unusual properties. We discuss these in the following sub-sections, concentrating on the detection with the highest outlier index. The light curves, timing analyses, spectra and spectral fit parameters of the other observations of each source are provided in the Appendix for completeness.

In all cases the \emph{XMM-Newton} data were reduced using the \texttt{Science Analysis Software} (\texttt{SAS}) v13.5 and the latest calibration files as of 2014 August 21, using the same method outlined in \citet{cal12}. X-ray spectral fitting was performed using \texttt{XSPEC} v12.8.1g \citep{arn96} over energies between 0.3 -- 10 keV, and the spectra were binned at 20 counts per bin to provide sufficient statistics for $\chi^2$ fitting. Photoelectric absorption was accounted for using the phabs model in \texttt{XSPEC} with the \citet{wil00} elemental abundances.

\subsection{3XMM J184430.9$-$024434: A Supergiant Fast X-ray Transient?}

3XMM J184430.9$-$024434 was observed twice with \emph{XMM-Newton} on the 15th and 16th of April 2010, and has two detections with the MOS2 camera in 3XMM (it fell off the chip in the pn and MOS1 exposures for both observations). The highest outlier measure of 1517 is in the first observation and it has a classifier margin of $-0.02$. It has a 3XMM 0.2--12 keV flux of $\sim8~\times~10^{-13}$ erg cm$^{-2}$ s$^{-1}$ and lies within the Galactic plane, $\sim$30$^\circ$ from the Galactic centre. No counterpart was found in NOMAD or the radio catalogs within the 3$\sigma$ positional errors, although there is a NOMAD source 2.8$\arcsec$ from the 3XMM position with B $\sim$ 19.1 mag, R $\sim$ 17.3 mag, J $\sim$ 14.8 mag, H $\sim$ 13.9 mag, and K $\sim$ 13.4 mag. 

\begin{figure}
\includegraphics[width=\columnwidth]{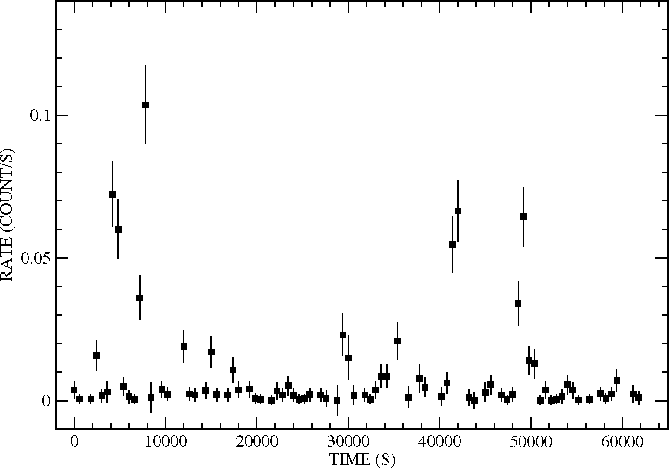}
\caption{EPIC MOS2 light curve (bin size =  600 s) of the candidate SFXT 3XMM J184430.9$-$024434 from observation 1.}\label{sfxt_lc}
\end{figure}

The 3XMM light curves show 5 distinct short, sharp flares in the first observation (Figure \ref{sfxt_lc}) and a single large flare in the second, reaching count rates of $\sim$0.1 count s$^{-1}$ before dropping back to zero. Such flares are reminiscent of supergiant fast X-ray transients (SFXTs), a rare sub-class of high mass X-ray binary (HMXB) of which $\sim$10 sources are currently known \citep[see][for a recent review]{sid14}. SFXTs are characterised by short duration X-ray flares (typically lasting $\sim10^2-10^3$ s) produced by transient accretion onto a compact object (typically a neutron star) from the wind of a blue supergiant companion \citep{sid14}. The dense wind environment produces high levels of photo electric absorption in their X-ray spectra and many SFXTs contain X-ray pulsars with spin periods of $\sim10-10^3$ s.

\begin{figure}
\includegraphics[width=\columnwidth]{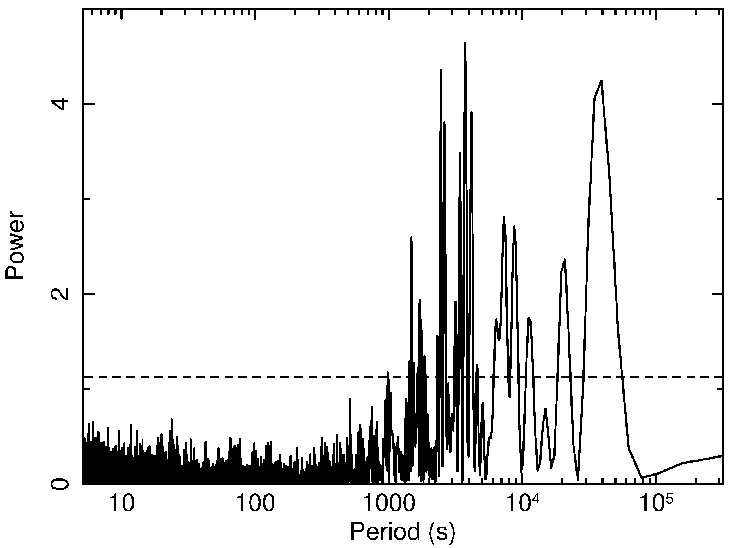}
\caption{Lomb-Scargle power spectrum of the MOS2 light curve (binned at the frame time of 2.6 s) of the candidate SFXT 3XMM J184430.9$-$024434 from observation 1. The 99$\%$ significance level is indicated by the dashed line.}\label{sfxt_ps}
\end{figure}

We extracted source and background light curves (from circular regions of radii 25$\arcsec$) binned at the frame time of 2.6 s, filtering out times with high background flaring. We corrected and background subtracted the light curves using the \texttt{SAS} task \texttt{epiclccorr} and applied a barycentric correction. We then searched for periodic variability using the \texttt{fasper} implementation of the Lomb-Scargle periodogram \citep{pre89}, and used Monte Carlo simulations to determine the 99$\%$ white noise significance levels \citep[e.g.][]{kon98}. No evidence of periodic modulation was found in the power spectra of either observation, though significant power was observed at low frequencies due to the flaring (see Figure \ref{sfxt_ps}).

\begin{figure}
\includegraphics[width=\columnwidth]{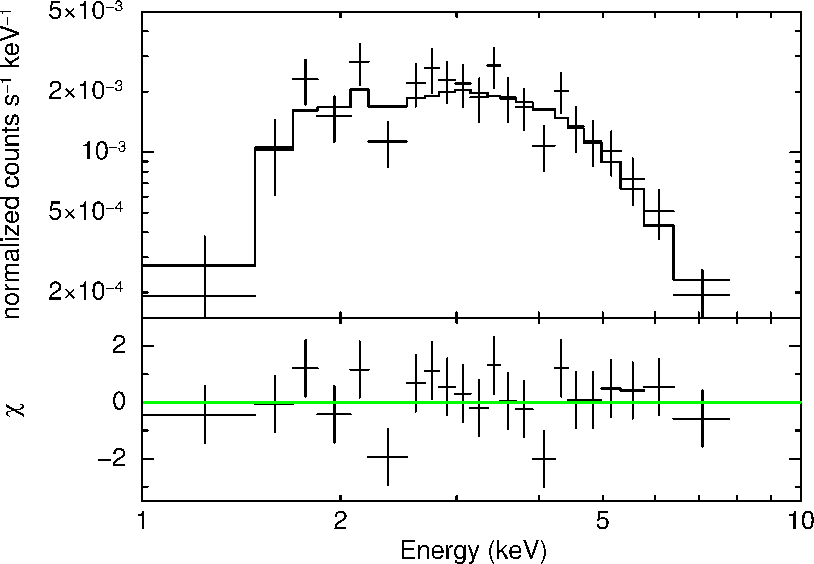}
\caption{EPIC MOS2 X-ray spectrum of the candidate SFXT 3XMM J184430.9$-$024434 fitted with an absorbed power law.}\label{sfxt_spec}
\end{figure}

We also extracted spectra from circular source (radius =  25$\arcsec$) and background (radius = 75$\arcsec$) regions and generated response and ancillary response files. We fitted the spectrum from the first observation with a number of simple models including a power law, black body, bremsstrahlung, and thermal plasma (all with photoelectric absorption components). We obtained the best fit with a power law model with high levels of photoelectric absorption ($\chi^2$/dof = 17.4/19). The spectrum of the second observation was well fitted by a similar model with $\chi^2$/dof = 5.6/11. Figure \ref{sfxt_spec} shows the spectrum from observation 1 fitted with an absorbed power law. Table \ref{sfxt_spec_prop} in the Appendix lists the spectral parameters for the best-fit models. The X-ray spectra and flare behavior are consistent with this source being a new member of the SFXT class.

\subsection{3XMM J181923.7$-$170616: A Slow X-ray Pulsar?}

3XMM J181923.7$-$170616 was observed three times with \emph{XMM-Newton} on the 7th of October 2006 (obsid = 0402470101), the 21st of March 2010 (obsid = 0604820101), and the 21st of March 2013 (obsid = 0693900101). Only the first two of these observations are in 3XMM, with a total of 6 EPIC detections. The highest outlier measure of 1782 is from the pn detection in the second observation and it has a classifier margin of $-0.10$. It has a 3XMM 0.2--12 keV flux of $5~\times~10^{-12}$ erg cm$^{-2}$ s$^{-1}$ and lies within the Galactic plane, $\sim$14$^\circ$ from the Galactic centre. It has a NOMAD counterpart with J $\sim$ 16.0 mag, H $\sim$ 14.0 mag, and K $\sim$ 13.5 mag but no match in the radio or galaxy catalogs. 

\begin{figure}
\includegraphics[width=\columnwidth]{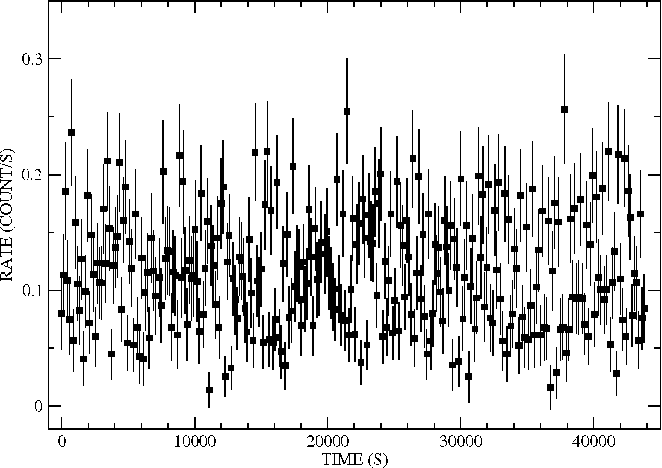}
\caption{EPIC pn light curve (bin size = 150 s) of the candidate slow pulsar 3XMM J181923.7$-$170616 from observation 2.}\label{psr_lc2}
\end{figure}

\begin{figure}
\includegraphics[width=\columnwidth]{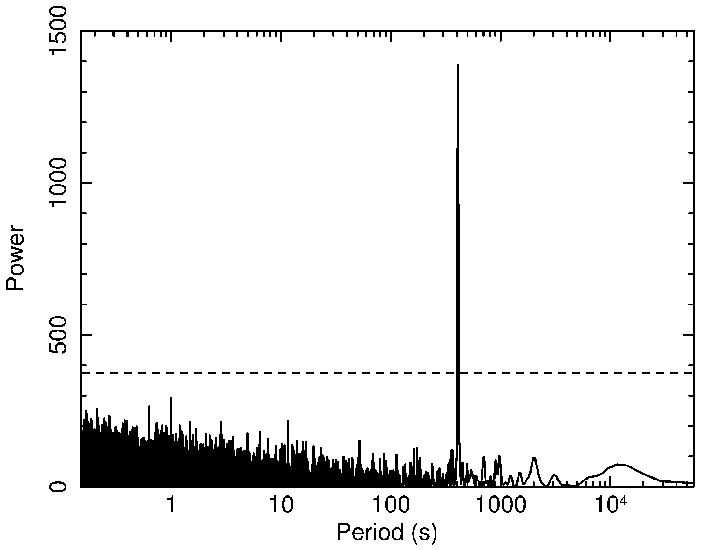}
\caption{Lomb-Scargle power spectrum of the pn light curve (binned at the frame time of 73.4 ms) of the candidate slow pulsar 3XMM J181923.7$-$170616 from observation 2. The 99$\%$ significance level is indicated by the dashed line.}\label{psr_ps2}
\end{figure}

\begin{figure}
\includegraphics[width=\columnwidth]{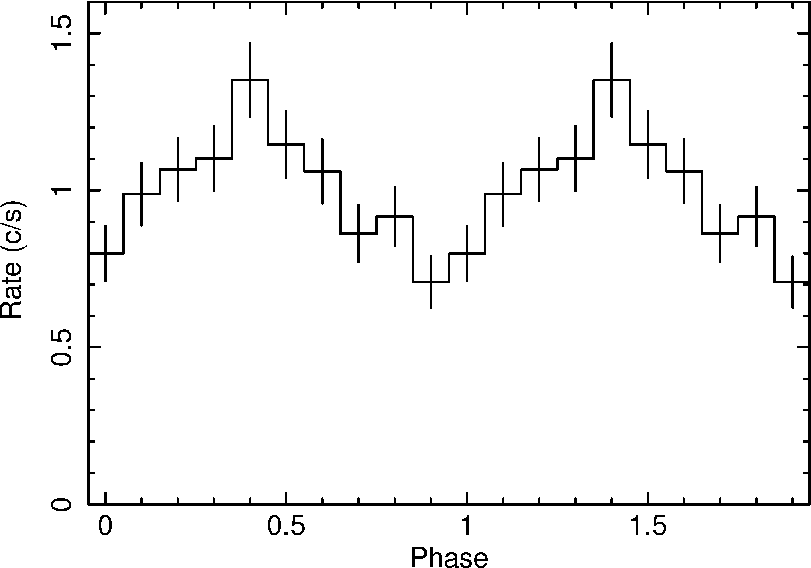}
\caption{EPIC pn light curve of the candidate slow pulsar 3XMM J181923.7$-$170616 from observation 1 folded over a period of 400 s.}\label{psr_fold1}
\end{figure}

The 3XMM light curves show no obvious features, though it is moderately variable with a fractional variability amplitude of $\sim$0.3 (Figure \ref{psr_lc2} and Figures \ref{psr_obs1} and \ref{psr_obs3} in the Appendix). We extracted source and background light curves (from circular regions of radii 30$\arcsec$) from the pn data binned at the frame time of 73.4 ms, filtering out times with high background flaring. We then corrected and background subtracted the light curves using the same method as described in \S5.1 and searched for periodic variability using the \texttt{fasper} implementation of the Lomb-Scargle periodogram. A significant peak was found in the power spectrum of the observation 2 pn light curve at a period of 400 s (Figure \ref{psr_ps2}). The profile of the light curve when folded over a period of 400 s (using the \texttt{efold} task in the \texttt{FTOOLS} software package) is roughly sinusoidal (see Figure \ref{psr_fold1}). The same periodic variability was also detected in the other two observations. This 400 s period was not detected in our automatic analysis of the 3XMM light curves that was used to generate the timing features. However, the generalised Lomb-Scargle method used to generate our timing features searched for periods only down to four times the bin width, insufficient to detect a period of $\sim$400 s given that the light curve was binned at 150 s. 

We extracted spectra from source (radius = 30$\arcsec$) and background (radius = 90$\arcsec$) regions and generated response and ancillary response files. We fitted the pn, MOS1, and MOS2 spectra from observation 2 simultaneously with simple absorbed power law, black body, bremsstrahlung, and thermal plasma models. An additional constant multiplicative component (frozen at 1 for the pn spectrum) was included to account for differences in the instrument responses. The best fit was obtained with the power law model ($\chi^2$/dof = 481.9/377), though the fit is not statistically acceptable and the residuals indicate that the model did not adequately represent the data above $\sim$6 keV. 

\begin{figure}
\includegraphics[width=\columnwidth]{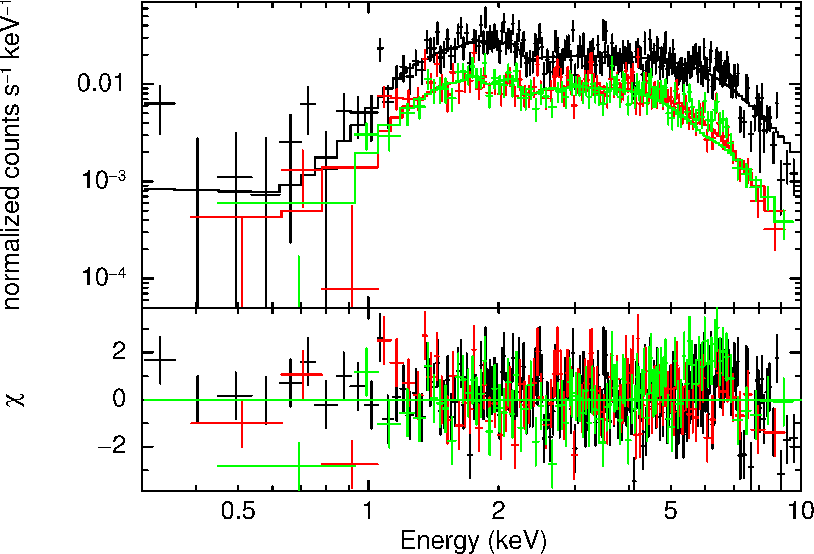}
\caption{EPIC X-ray spectra (black = pn, red = MOS1, green = MOS2) from observation 2 of the candidate slow pulsar 3XMM J181923.7$-$170616 fitted with an absorbed power law model.}\label{psr_spec}
\end{figure}

The addition of a high energy exponential cut-off improved the fit ($\chi^2$/dof = 454.5/375), while adding an additional gaussian line at $\sim$6.7 keV (representing Helium-like iron emission) improved it further ($\chi^2$/dof = 431.5/372). To test whether the addition of these model components is statistically justifiable we calculated the Bayesian information criterion (BIC) for each model. We found that the BIC was lowest for the simple absorbed power law model, indicating that adding the high energy cut-off and the Gaussian emission line are not statistically justified. The poor fit residuals could, however, be indicative of spectral variability within the observation. Further analysis is needed to better constrain the nature of the X-ray emission. Figure \ref{psr_spec} shows the spectra of observation 2 fitted with the simple absorbed power law model, while Table \ref{psr_spec_prop} presents the parameters of this fit. The spectra from observations 1 and 3 were also well described by an absorbed power law. 

The $\sim$400 s period could represent the spin of a compact object, either a white dwarf (i.e. a CV) or neutron star (i.e. an XRB). However, the spectra (in particular the moderately high absorption) are more consistent with a HMXB, a number of which are known to contain slowly spinning neutron stars with periods of hundreds of seconds \citep[e.g.][]{ikh14}.

\subsection{3XMM J181355.6$-$324237: An Eclipsing Binary?}

3XMM J181355.6$-$324237 was observed three times by \emph{XMM-Newton} on the 15th, 17th, and 19th of September 2009 (Obsids: 0604860201, 0604860301, and 0604860401). In all three observations the MOS cameras were in large window mode meaning that the source fell off the chip in all observations. There are therefore only 3 pn detections in 3XMM. The highest outlier measure of 523 is from the first observation, and the classification margin is $\sim$0. The 3XMM 0.2--12 keV flux is $\sim4~\times~10^{-13}$ erg cm$^{-2}$ s$^{-1}$ in the first two observations, but drops to $\sim3~\times~10^{-14}$ erg cm$^{-2}$ s$^{-1}$ in the third. It is located towards the Galactic centre but $\sim$7$^\circ$ degrees below the plane, and is coincident with the Cepheid variable star V2719 Sgr. It has no match in NOMAD within the 3$\sigma$ positional errors, but there are two NOMAD sources $\sim$1.5$\arcsec$ away. Inspection of the \emph{XMM-Newton} optical monitor images found a single source coincident with the X-ray position, indicating that there are likely some issues with the NOMAD astrometry and/or that the star has a high proper motion. The NOMAD source 0572-1005906 has B $\sim$ 12.6 mag, V $\sim$ 16.3 mag, R $\sim$ 17.3 mag, J $\sim$ 15.5 mag, H $\sim$ 14.9 mag, and K $\sim$ 14.9 mag.

\begin{figure}
\includegraphics[width=\columnwidth]{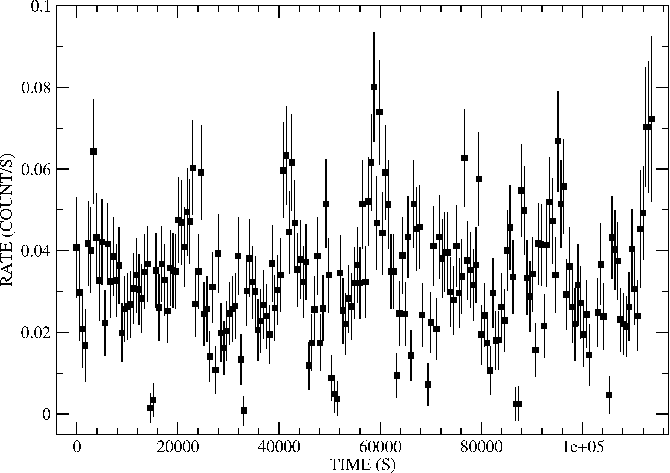}
\caption{EPIC pn light curve (bin size = 560 s) of the candidate eclipsing binary 3XMM J181355.6$-$324237 from observation 1.}\label{ceph_lc1}
\end{figure}

\begin{figure}
\includegraphics[width=\columnwidth]{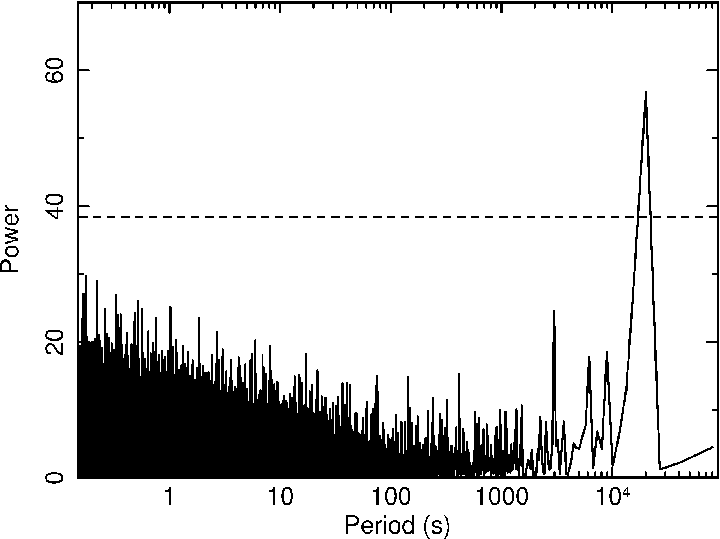}
\caption{Lomb-Scargle power spectrum of the EPIC pn light curve (binned at the frame time of 73.4 ms) of the candidate eclipsing binary 3XMM J181355.6$-$324237 from observation 1. The 99$\%$ significance level is indicated by the dashed line.}\label{ceph_ps1}
\end{figure}

\begin{figure}
\includegraphics[width=\columnwidth]{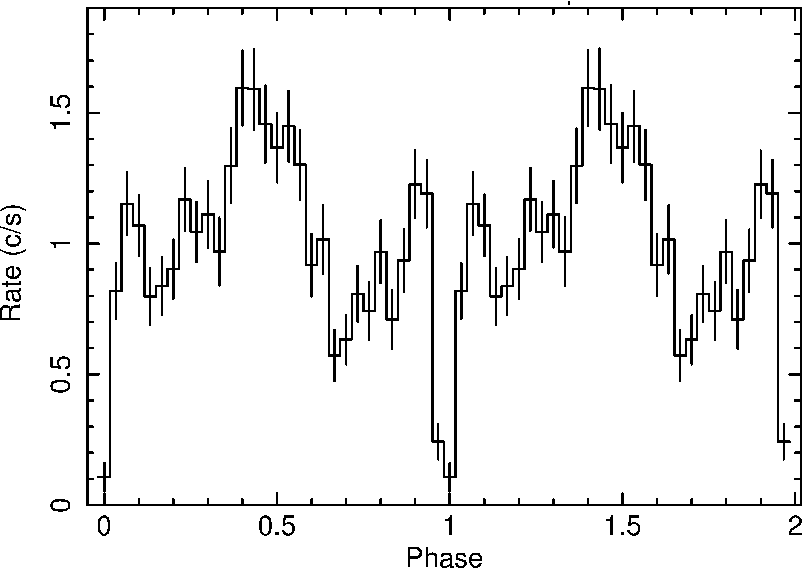}
\caption{EPIC pn light curve of the candidate eclipsing binary 3XMM J181355.6$-$324237 from observation 1 folded over a period of 18 ks.}\label{ceph_fold1}
\end{figure}

The 3XMM light curves show clear evidence for periodic dips over a period of $\sim$20 ks in the first two observations (see Figure \ref{ceph_lc1}) where the count rate drops to zero for $\sim$1 ks. No evidence of dips is seen in the third observation, though this is likely due to the poorer S/N resulting from the significant drop in flux. Our automatic timing analysis routine indicated the presence of significant periodic variability with a period of 18,523 s with a false alarm probability of $\sim$10$^{-14}$. To test for higher frequency variability, we extracted source and background light curves from all three pn exposures at the frame time of 73.4 ms (from circular regions of radii 15$\arcsec$ for observations 1 and 2, and 12$\arcsec$ for observation 3), correcting them as described above. A strong peak was detected at a period of $\sim$18 ks in the power spectra of the observation 1 (Figure \ref{ceph_ps1}) and 2 light curves, but no evidence was found for periodic variability at any other frequency. No periodic variability was detected at any frequency in the observation 3 light curve. The profile of the observation 1 light curve folded over a period of 18 ks is shown in Figure \ref{ceph_fold1}. A clear dip is seen out of phase with the light curve maximum. Such dips are typical of eclipsing binary systems such as CVs and XRBs.

\begin{figure}
\includegraphics[width=\columnwidth]{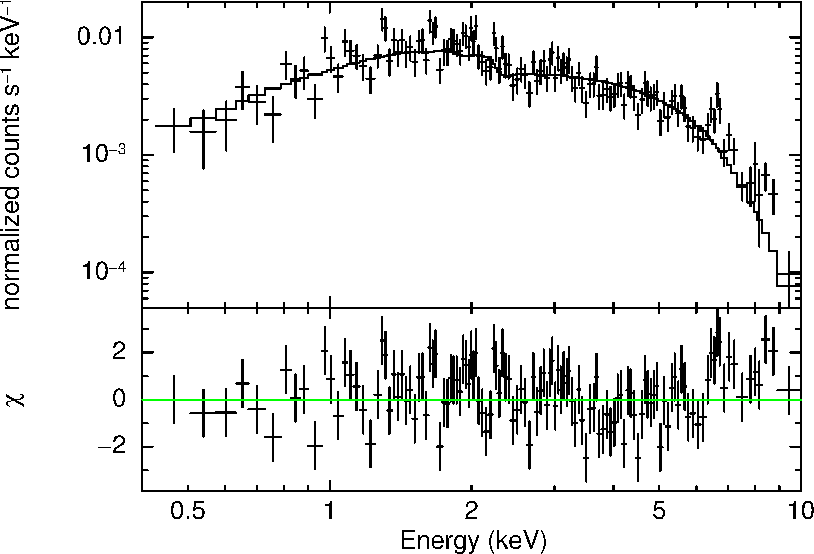}
\caption{EPIC pn spectrum of the candidate eclipsing binary 3XMM J181355.6$-$324237 from observation 1 fitted with an absorbed black body model.}\label{ceph_spec1}
\end{figure}

We extracted spectra using the same source regions as for the light curves but with background regions of radii 45$\arcsec$ for observations 1 and 2 and 36$\arcsec$ for observation 3. We fitted the spectrum from observation 1 with the same simple models as before, obtaining the best fits with absorbed black body ($\chi^2$/dof = 189.3/127) and power law ($\chi^2$/dof = 193.5/127) models. The residuals indicated problems at high energies for the black body fit and at low energies for the power law model. Fitting the spectrum with a combined power law plus low temperature black body model obtained a better fit ($\chi^2$/dof = 162.4/125), while the addition of a Gaussian emission line at $\sim$6.7 keV improved the fit even further ($\chi^2$/dof = 145.4/121). We calculated the BIC for each model, finding that it was lowest for the simple absorbed black body model, indicating that the power law and the Gaussian emission line are not statistically required. As with our candidate slow pulsar, further analysis is needed to better constrain the nature of the X-ray emission. The spectrum of observation 1 fitted with the simple absorbed black body model is shown in Figure \ref{ceph_spec1}, while Table \ref{ceph_spec_prop} presents the spectral parameters of this fit. The spectrum in observation 2 is very similar to that of observation 1. However, in observation 3 the spectrum changed significantly in shape such that a simple absorbed black body model was no longer an acceptable fit ($\chi^2$/dof = 57.9/22) and an absorbed power law provides a much better approximation of the data ($\chi^2$/dof = 25.1/22). 

The presence of periodic sharp dips in the light curves is consistent with an eclipsing binary system, implying an orbital period of $\sim$18 ks and an orbital radius of $\sim$10$^{6}$ km (assuming a combined mass of $\sim$1 -- 10 M$_\odot$ for both stars). However, a Cepheid would not fit inside such a tight orbit as Galactic Cepheids have been found to have radii $\sim$10$^{7}$ -- 10$^{8}$ km \citep[e.g.][]{gie98}. It is possible that the Cepheid classification for V2719 Sgr was incorrect, or that it is aligned by chance with 3XMM J181355.6$-$324237. Alternatively, if V2719 Sgr truly is a Cepheid and associated with this 3XMM source, it could be a hierarchical triple system with a compact binary orbited by a Cepheid in a wider orbit. The X-ray spectra are most consistent with an XRB, however we cannot rule out a CV on this data alone. Regardless, 3XMM J181355.6$-$324237 appears likely to be a new eclipsing binary system.

\section{Summary \& Conclusions}

We have applied the RF machine learning algorithm to automatically classify 2,876 variable X-ray sources in the 3XMM catalog. We obtained a classification accuracy of $\sim92\%$ when classifying by source class. We also tested the application of RF for quality control classification (i.e. identifying spurious sources), obtaining an accuracy of $\sim95\%$. Manual investigation of classified unknown sources found that 96$\%$ of a sample of 200 randomly selected sources were correctly classified as either spurious or non-spurious. We also manually tested the accuracy of the classification by source class for a random sample of sources with P(Spur) $\leqslant$ 30\%, finding that 92$\%$ of a sample of 101 sources with previous identifications in the literature were correctly classified. As with our previous work \citep{lo14}, we found that RF had trouble classifying sources belonging to classes that were not adequately sampled in the training set (e.g. GRBs). Regardless, selecting sources with P(Max) $\geqslant$ 60\% and P(Spur) $\leqslant$ 30\% should provide a clean sample of real sources with predominantly correct classifications.

We also investigated a sample of 144 anomalous good sources with outlier measures $\geqslant$ 400 or classifier margins $\leqslant~-0.3$. We identified a number of truly rare sources that were not represented in the training set (e.g. isolated neutron stars, magnetars, unusual AGN etc.), validating the effectiveness of the RF classifier for identifying members of rare outlier source populations. We also identified a number of sources with unreliable X-ray parameters and/or astrometry that were detected in problematic mosaic mode observations. In addition, we identified three previously unstudied sources that appear to be truly unique objects, including a new candidate SFXT, a new candidate 400 s slow pulsar, and an eclipsing compact binary system with a 5 hr orbital period that may be the inner binary of a hierarchical triple system containing a Cepheid variable. Additional work beyond the scope of this paper is underway to further investigate these unique objects. 

\acknowledgments

We thank the anonymous referee for their careful review of this paper, Ben Chan for providing us with the NED catalog of galaxies, and Tom Maccarone and Simon Vaughan for useful discussions. SAF and TM acknowledge the support of the Australian Research Council through Discovery Project DP110102034. This research was conducted by the Australian Research Council Centre of Excellence for All-sky Astrophysics (CAASTRO), through project number CE110001020. Based on observations from XMM-Newton, an ESA science mission with instruments and contributions directly funded by ESA Member States and NASA. This work made use of the 3XMM Serendipitous Source Catalog, constructed by the XMM-Newton Survey Science Centre on behalf of ESA. This research has also made use of the NASA/IPAC Extragalactic Database (NED) which is operated by the Jet Propulsion Laboratory, California Institute of Technology, under contract with the National Aeronautics and Space Administration, and the SIMBAD database, operated at CDS, Strasbourg, France. 


%% in parentheses, after the keyword, but they will not be verified.

{\it Facilities:} \facility{XMM-Newton}.



\appendix

\section{Appendix}

Here we provide light curves, Lomb-Scargle power spectra, and spectra for the additional \emph{XMM-Newton} observations for the three outliers sources discussed in \S5. We also provide tables giving the best-fit spectral models for all EPIC spectra for these sources.

% SFXT candidate
\subsection{The Candidate SFXT 3XMM J184430.9$-$024434}

\begin{figure}
\includegraphics[height=19cm]{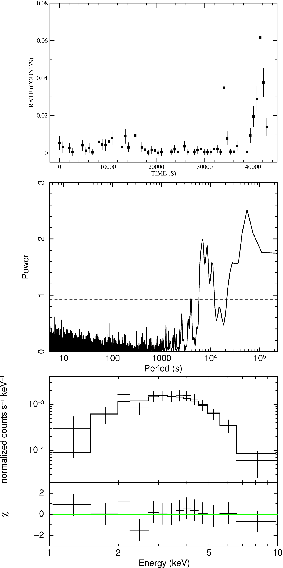}
\caption{\emph{Top:} EPIC MOS2 light curve (bin size = 680 s) of the candidate SFXT 3XMM J184430.9$-$024434 from observation 2. \emph{Middle:} Lomb-Scargle power spectrum of the MOS2 light curve (binned at the frame time of 2.6 s) of the candidate SFXT 3XMM J184430.9$-$024434 from observation 2. The 99$\%$ significance level is indicated by the dashed line. \emph{Bottom:} EPIC MOS2 X-ray spectrum of the candidate SFXT 3XMM J184430.9$-$024434 from observation 2 fitted with an absorbed power law.}\label{sfxt_obs2}
\end{figure}

\begin{table}
\begin{center}
\caption{Parameters of the best-fit models fitted to the EPIC spectra of the candidate SFXT 3XMM J184430.9$-$024434.\label{sfxt_spec_prop}}
\begin{tabular}{lccc}
\tableline\tableline 
Parameter & Observation 1 & Observation 2 & Units\\
\tableline
$nH$ & 5$^{+2}_{-1}$ & 7$^{+5}_{-3}$ & 10$^{22}$ atom cm$^{-2}$\\
$\Gamma$ & 1.5$^{+0.6}_{-0.5}$ & 1.8$^{+0.9}_{-0.7}$ & \nodata \\
Flux$^a$ & 7$\pm$1 & 6$\pm$1  & 10$^{-13}$ erg cm$^{-2}$ s$^{-1}$\\
$\chi^2$/dof & 17.4/19 & 5.6/11& \nodata \\
\tableline
\multicolumn{4}{l}{$^a$ Absorbed flux in the 0.2--10 keV band.}\\
\end{tabular}
\end{center}
\end{table}

\clearpage

% Slow pulsar candidate 
\subsection{The Candidate Slow Pulsar 3XMM J181923.7$-$170616}

\begin{figure}
\includegraphics[height=19cm]{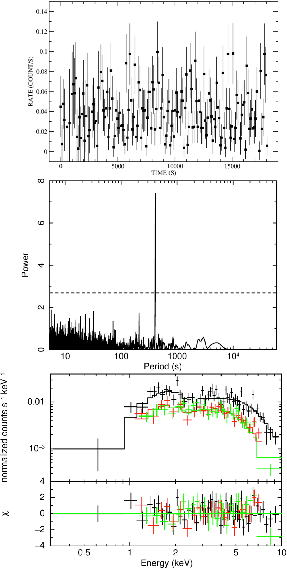}
\caption{\emph{Top:} EPIC MOS1 light curve (bin size = 110 s) of the candidate slow pulsar 3XMM J181923.7$-$170616 from observation 1. \emph{Middle:} Lomb-Scargle power spectrum of the MOS1 light curve (binned at the frame time of 2.6 s) of the candidate slow pulsar 3XMM J181923.7$-$170616 from observation 1. The 99$\%$ significance level is indicated by the dashed line. \emph{Bottom:} EPIC X-ray spectra (black = pn, red = MOS1, green = MOS2) of the candidate slow pulsar 3XMM J181923.7$-$170616 from observation 1 fitted with an absorbed power law model.}\label{psr_obs1}
\end{figure}

\begin{figure}
\includegraphics[height=19cm]{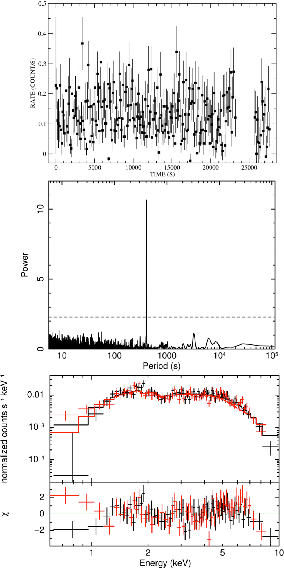}
\caption{\emph{Top:} EPIC MOS1 light curve (bin size = 110 s) of the candidate slow pulsar 3XMM J181923.7$-$170616 from observation 3. \emph{Middle:} Lomb-Scargle power spectrum of the MOS1 light curve (binned at the frame time of 2.6 s) of the candidate slow pulsar 3XMM J181923.7$-$170616 from observation 3. The 99$\%$ significance level is indicated by the dashed line. \emph{Bottom:} EPIC X-ray spectra (black = MOS1, red = MOS2) of the candidate slow pulsar 3XMM J181923.7$-$170616 from observation 1 fitted with an absorbed power law model.}\label{psr_obs3}
\end{figure}

\begin{table}
\begin{center}
\caption{Parameters of the best-fit models fitted to the EPIC spectra of the candidate slow pulsar 3XMM J181923.7$-$170616.\label{psr_spec_prop}}
\begin{tabular}{lcccc}
\tableline\tableline 
Parameter & Observation 1 & Observation 2 & Observation 3 & Units\\
\tableline
$nH$ & 1.2$^{+0.4}_{-0.3}$ & 1.5$\pm$0.2 & 1.2$\pm$0.2  &  10$^{22}$ atom cm$^{-2}$\\
$\Gamma$ & 0.4$\pm$0.1 & 0.56$\pm$0.07 & 0.4$\pm$0.1  &  \nodata \\
Flux$^a$ & 3.2$^{+0.2}_{-0.1}$ & 3.35$^{+0.10}_{-0.09}$ & 4.2$^{+0.3}_{-0.1}$  &  10$^{-12}$ erg cm$^{-2}$ s$^{-1}$\\
$\chi^2$/dof & 110.0/94 & 481.9/377 & 157.2/127  & \nodata \\
\tableline
\multicolumn{5}{l}{$^a$Absorbed flux in the 0.2--10 keV band.}\\
\end{tabular}
\end{center}
\end{table}

\clearpage

% Eclipsing LMXB candidate
\subsection{The Candidate Eclipsing Binary 3XMM J181355.6$-$324237}

\begin{figure}
\includegraphics[height=19cm]{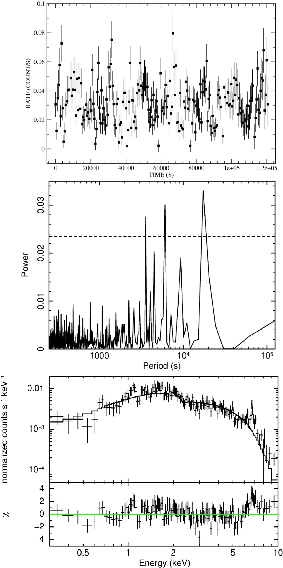}
\caption{\emph{Top:} EPIC pn light curve of the candidate eclipsing binary 3XMM J181355.6$-$324237 from observation 2. \emph{Middle:} Lomb-Scargle power spectrum of the EPIC pn light curve (binned at 100 s) of the candidate eclipsing binary 3XMM J181355.6$-$324237 from observation 2. The 99$\%$ significance level is indicated by the dashed line. \emph{Bottom:} EPIC pn spectrum of the candidate eclipsing binary 3XMM J181355.6$-$324237 from observation 2 fitted with an absorbed black body model.}\label{ceph_obs2}
\end{figure}

\begin{figure}
\includegraphics[height=19cm]{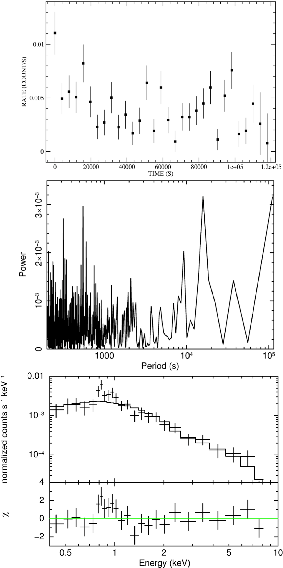}
\caption{\emph{Top:} EPIC pn light curve of the candidate eclipsing binary 3XMM J181355.6$-$324237 from observation 3. \emph{Middle:} Lomb-Scargle power spectrum of the EPIC pn light curve (binned at 100 s) of the candidate eclipsing binary 3XMM J181355.6$-$324237 from observation 3. The 99$\%$ significance level is greater than 5 $\times$ 10$^{-3}$ and is thus not shown. \emph{Bottom:} EPIC pn spectrum of the candidate eclipsing binary 3XMM J181355.6$-$324237 from observation 3 fitted with an absorbed power law.}\label{ceph_obs2}
\end{figure}

\begin{table}
\begin{center}
\caption{Parameters of the best-fit models fitted to the EPIC spectra of the candidate eclipsing binary 3XMM J181355.6$-$324237.\label{ceph_spec_prop}}
\begin{tabular}{lcccc}
\tableline\tableline 
Parameter & Observation 1 & Observation 2 & Observation 3 & Units\\
\tableline
$nH$ &  0.09$^{+0.06}_{-0.03}$&  0.002$^{+0.050}_{-0.002}$ & 0.15$^{+0.10}_{-0.09}$ & 10$^{22}$ atom cm$^{-2}$\\
$kT$ & 1.34$\pm$0.06 & 1.27$^{+0.05}_{-0.06}$ &  \nodata & \nodata \\ 
$\Gamma$ & \nodata & \nodata &  1.9$^{+0.4}_{-0.3}$ & \nodata \\ 
Flux$^a$ & 2.8$\pm$0.2 & 2.5$\pm$0.1 & 0.22$^{+0.05}_{-0.04}$ & 10$^{-13}$ erg cm$^{-2}$ s$^{-1}$\\
$\chi^2$/dof & 181.9/126 & 200.7/135 & 25.1/22 & \nodata \\
\tableline
\multicolumn{5}{l}{$^a$Absorbed flux in the 0.2--10 keV band.}\\
\end{tabular}
\end{center}
\end{table}




\clearpage

\end{document}